\documentclass[12pt]{article}
\title{ Spin-dependent interaction  in the deconfined phase of QCD}
\author{Yu.A.Simonov\\
State Research
Center\\Institute of Theoretical and Experimental Physics, \\
Moscow, 117218 Russia}

\newcommand{\beq}{\begin{eqnarray}}
 \newcommand{\eeq}{\end{eqnarray}}
\newcommand{\be}{\begin{equation}}
 \newcommand{\ee}{\end{equation}}

\def\fun#1#2{\lower3.6pt\vbox{\baselineskip0pt\lineskip.9pt
\ialign{$\mathsurround=0pt#1\hfil ##\hfil$\crcr#2\crcr\sim\crcr}}}

\newcommand{{\SD}}{\rm SD}

\newcommand{\vesig}{\mbox{\boldmath${\rm \sigma}$}}

\newcommand{\veS}{\mbox{\boldmath${\rm S}$}}
\newcommand{\veL}{\mbox{\boldmath${\rm L}$}}
\newcommand{\veR}{\mbox{\boldmath${\rm R}$}}

\newcommand{\veB}{\mbox{\boldmath${\rm B}$}}

\newcommand{\veE}{\mbox{\boldmath${\rm E}$}}

\newcommand{\llan}{\langle\langle}
\newcommand{\rran}{\rangle\rangle}
\newcommand{\lan}{\langle}
\newcommand{\ran}{\rangle}

\begin{document}
\maketitle
\begin{abstract}
 Spin-dependent deconfined  interaction in the $Q\bar
Q$  system is derived from the field correlators known from
lattice and analytic calculations. As a result hyperfine splitting
is found numerically for charmonium, bottomonium and strangeonium
in the range $T_c\leq T \leq 2T_c$. Spin-orbit interaction due to
magnetic correlators (the Thomas term) is able to produce numerous
$Q\bar Q$ bound states with accumulation point at $M=m_Q+m_{\bar
Q}$. Possible influence of these effects on the thermodynamics of
quark-gluon plasma is discussed.

\end{abstract}
\hfill {\it In honor of A.Di Giacomo on his seventieth birthday.}

\section{Introduction}

 The Spin-Dependent Interaction (SDI) in  the $Q\bar Q$ system is
 being studied for the last  three decades \cite{1}-\cite{12} for heavy
 qurkonia in the confinement phase. The first estimates of SDI \cite{1,5} essentially exploited perturbative
 expansion and nonperturbative SDI   appeared in the form of the
 Thomas term \cite{2}. In the general approach  \cite{3,4} using
 $1/M_q$ expansion for heavy quarks, four spin-dependent static
 potentials $V_i(r), i=1,2,3,4 $ are expressed through connected
 field correlators.

 In this approach perturbative expansion of correlators
 reproduces perturbative part of $V_i(r)$ while the
 nonperturbative part can be obtained only from lattice
 simulations, done by several groups \cite{6,7,8}.

 In the Field-Correlator Method (FCM) \cite{9} (see \cite{10} for
 a review) the SDI  potentials have been obtained using $1/M_q$
 expansion in \cite{11,12}, in the  form containing only Gaussian
 correlators $D(x)$, $D_1(x)$, and allowing for  immediate check
 of Gromes relation \cite{4}. The nonperturbative component of
 $V_i(r)$ is calculated directly through $D(x), D_1(x)$ and the
 Thomas term naturally appears as asymptotics at large distances.
Extensive and thorough lattice calculations of correlators $D(x),
D_1(x)$ done  by the Pisa group \cite{13} have made it possible to
determine numerically all 4 potentials $V_i$.

The first comparison of resulting spin splittings of charmonium
and bottomonium levels   was  done in \cite{14} and has   shown a
good agreement of lattice-based potentials, with   experiment. At
the same time spin-orbit splittings were found to be sensitive to
the  behaviour of field correlators  $D(x), D_1(x)$ at small $x$.

 In \cite{15} a detailed analysis was done of SD potentials,
 obtained from field correlators  in \cite{11}, as compared  to
 the definitions of Eichten-Feinberg-Gromes (EFG) \cite{3,4}. In
 the latter case SD potentials $V_i$ are not shown to satisfy
 Gromes relation \cite{4}, and analysis of \cite{15} has
 demonstrated the difficulties arising in this respect for  the lattice-defined $V_i $ \cite{6,7}.

 It was understood later \cite{16}, that the same set of SDI
 potentials $V_i(r)$ can be obtained without the   $1/M_q$
 expansion, using cluster  expansion of gluonic fields for quarks of arbitrary mass
 $m_q$ and in this case instead of quark masses $m_q$, there
 appear in $V_i(r)$ the so-called einbein masses $\mu_q$, with
 effective average values $\lan \mu_q\ran$ playing the role of
 constituent masses calculated through $m_q$ and string tension
 $\sigma$.

 This approach works well in the confinement phase for meson \cite{16} and
 baryon \cite{17} states of light and heavy quarks, when SDI can be
 considered as corrections, see e.g. the analysis \cite{18}  for
 light scalar mesons, and \cite{19} for heavy quarkonia.

 The inclusion of chiral dynamics (where SDI is crucial) needs
 modification in the formalism, and instead of SDI potentials one
 considers  in this case the effective Lagrangians, as in \cite{20}, and the huge mass splitting between $\rho$ and $\pi$ is
 obtained as the chiral nonlinear amplification of the standard
 hyperfine SDI.

 So far so good for the zero temperature QCD. For $T>0$ there
 appears a first change in SDI namely one must distinguish
 colorelectric and colormagnetic correlators,  contributing to
 $V_i(r)$, since instead of two  Gaussian correlators $D(x),
 D_1(x)$ one has for $T>0$ four independent correlators: $D^E(x),
 D^H(x), D_1^E(x), D_1^H(x)$ and it was agrued in \cite{21,22,23} that only one of them vanishes for $T>T_c$.
These correlators have been found numerically on the lattice
 \cite{24}, and the original expectation was confirmed, that the
 correlators do not change significantly for $T\ll M_d$, where
 $M_d$ is the dilaton mass of the order of $0^{++}$ glueball mass,
 $M_d\sim 1\div 1.5$ GeV.
Moreover it was shown in \cite{24} that indeed only one of
correlators, $D^E(x)$ vanishes  exactly at $T=T_c$.

Already at this stage one can show that all four SDI potentials
are expressed through magnetic correlators, and only one SDI term
is due to the  colorelectric correlator, contributing to
spin-orbit force.

The situation becomes even more interesting for $T>T_c$, where the
main colorelectric correlator $D^E(x)$, ensuring confinement,
disappears but all other correlators stay intact.

In this case all four SDI potentials should  change only  slightly
above $T_c$ and ensure the spin splitting of levels of the same
type as it was below $T_c$. Of special interest is the Thomas
term, which is  doubled in magnitude (due to disappearance of
colorelectric static force $\varepsilon (r)$) and  is dominant at
large distances. It is argued below in the paper, that the
spin-orbit Thomas term can create  infinite number of bound $Q\bar
Q$ states with the accumulation point near $2M_Q$.

\section{Spin-dependent interaction in  the $q\bar q$ system}

For the quark-antiquark Green's function ( of both light and heavy
quarks) one can use the Fock-Feynman-Schwinger Representation
(FSR) \cite{25} with the kernel containing vacuum fields and quark
spin operators in the form \cite{11,16}
$$ \lan W_{\sigma} (x,y) \ran = \lan \exp ig \int
d\pi_{\mu\nu} (z) F_{\mu\nu} (z)\ran=$$ \be= \exp
\sum^\infty_{n=1} \frac{(ig)^n}{n!} \int d\pi(1)... \int d\pi (n)
\llan F(1)...F(n)\rran,\label{1}\ee where \be d\pi_{\mu\nu} (z)= d
s_{\mu\nu} (z) -i \sigma_{\mu\nu} d\tau\label{2}\ee Here
$ds_{\mu\nu}$ is the surface element, while $d\tau$ is  the proper
time integration, which can be connected to usual time $t$ via
$d\tau_i=\frac{dt_i}{2\mu_i}$ \cite{16}. The spin-spin interaction
can be obtained  from (\ref{1}), (\ref{2}) keeping only quadratic
(Gaussian) correlators $ \lan F F\ran$, \be \exp \left\{ -\frac12
\int^{s_1}_0 d\tau_1 \int^{s_2}_0 d\tau_2 g^2 \left \lan\left (
\begin{array}{ll}\vesig^{(1)} \veB& \vesig^{(1)}
\veE\\\vesig^{(1)} \veE& \vesig^{(1)}
\veB\end{array}\right)_{z(\tau_1)} \left ( \begin{array}{ll}
\vesig^{(2)} \veB& \vesig^{(2)} \veE\\\vesig^{(2)} \veE&
\vesig^{(2)} \veB\end{array}\right)_{z(\tau_2 )}\right \ran
\right\},\label{3}\ee and spin-orbit interaction arises in
(\ref{1}) from the products $\lan\sigma_{\mu\nu} F_{\mu\nu}
ds_{\rho\lambda} F_{\rho\lambda}\ran$ \cite{11,16}.

It is clear, that the resulting SDI will be of matrix form
$(2\times 2)\times (2\times 2)$ ( not accounting for  Pauli
matrices). If  one keeps only diagonal terms in $\sigma_{\mu\nu}
F_{\mu\nu} $ (as the leading terms for large $\mu_i\approx M$) one
can write for the SDI the representation of the Eichten-Feinerg
form \cite{26}

$$ V_{SD}^{(diag)}(R)=(\frac{\vec{\sigma}_1\vec L_1}{4\mu_1^2}-
\frac{\vec{\sigma}_2\vec
L_2}{4\mu_2^2})(\frac{1}{R}\frac{d\varepsilon}{dR}+\frac{2dV_1(R)}{RdR})+
$$
\be + \frac{\vec{\sigma}_2\vec L_1- \vec{\sigma}_1\vec
L_2}{2\mu_1\mu_2}\frac{1}{R}\frac{dV_2}{dR}+\frac{
\vec{\sigma}_1\vec {\sigma}_2V_4(R)}{12\mu_1\mu_2}+\frac{(3
\vec{\sigma}_1 \vec R\vec{\sigma}_2 \vec R-\vec{\sigma}_1
\vec{\sigma}_2 R^2)V_3} {12\mu_1\mu_2 R^2}. \label{4} \ee

At this point one should note that the term with
$\frac{d\varepsilon}{dR}$ in (\ref{4})
 was obtained from the diagonal part of the matrix  $(m-\hat D)
 \sigma_{\mu\nu} F_{\mu\nu}$, namely as product  $ i\sigma_k
 D_k\cdot \sigma_i E_i$, see \cite{11,16} for details of
 derivation, while  all other potentials $V_i, i=1,2,3,4$ are
 proportional to correlators $\lan H_i \Phi H_k \Phi\ran$. One can
 relate correlators of colorelectric and colormagnetic fields to
 $D^E, D^E_1, D^H, D_1^H$  as follow (see \cite{11}, \cite{12})
\be
  g^2\lan E_i(\vec{x}_1,\tau_1)\Phi
 E_j(\vec x_2,\tau_2)\Phi^+\ran = N_c\left [\delta_{ij} \left (D^E(\lambda,
\nu)+D_1^E+\vec u^2\frac{\partial D_1^E}{\partial \vec u^2}\right
)+u_iu_j\frac{\partial D_1^E}{\partial u^2}\right ]
 \label{5}
  \ee

 \be
 g^2\lan H_i(x)\Phi H_j(y)\Phi^+\ran=N_c
 \left [\delta_{ij}\left (D^H(\lambda,\nu)+D_1^H+\vec u^2\frac{\partial
 D_1^H}{\partial\vec u^2}\right )-u_iu_j\frac{\partial D_1^H}{\partial\vec u^2}\right ]
 \label{6}
 \ee
 \be
 g^2\lan E_i(x)\Phi H_j(y)\Phi^+\ran =-N_c e_{ijn} u_n
 u_4\frac{\partial D_1^{EH}}{\partial\vec
 u^2})
 \label{7}
  \ee
As a result one obtains the following connection between SD
potentials and correlators.

\be
\frac{1}{R}\frac{dV_1}{dR}=-\int^{\infty}_{-\infty}d\nu\int^R_0
\frac{d\lambda}{R} \left (1-\frac{\lambda}{R}\right
)D^H(\lambda,\nu) \label{8} \ee \be
\frac{1}{R}\frac{dV_2}{dR}=\int^{\infty}_{-\infty}d\nu\int^R_0
\frac{\lambda d\lambda}{R^2} \left
[D^H(\lambda,\nu)+D_1^H(\lambda,\nu)+\lambda^2\frac{\partial
D_1^H}{\partial\lambda^2}\right ] \label{9} \ee \be
V_3=-\int^{\infty}_{-\infty} d\nu R^2\frac{\partial
D_1^H(R,\nu)}{\partial R^2} \label{10} \ee \be
V_4=\int^{\infty}_{-\infty}d\nu \left
(3D^H(R,\nu)+3D_1^H(R,\nu)+2R^2\frac{\partial D_1^H}{\partial
R^2}\right ) \label{11} \ee

 \be \frac{1}{R}\frac{d\varepsilon
(R)}{dR}=\int^{\infty}_{-\infty}d\nu\int^R_0 \frac{ d\lambda}{R}
\left
[D^E(\lambda,\nu)+D_1^E(\lambda,\nu)+(\lambda^2+\nu^2)\frac{\partial
D_1^E}{\partial\nu^2}\right ] \label{12} \ee

One can check, that the Gromes relation \cite{4} acquires the form
\be V'_1 (R) + \varepsilon'(R) -V'_2(R)=\int^\infty_{-\infty} d\nu
\left [ \int^R_0 d\lambda (D^E(\lambda, \nu)-D^H(\lambda,
\nu))+\frac12 R(D_1^E(R) -D_1^H(R))\right].\label{13}\ee

For $T=0$, when $D^E=D^H, D_1^E=D_1^H$, the Gromes relations are
satisfied identically, however for $T>0$ electric and magnetic
correlators are certainly different and Gromes relation is
violated, as one could tell beforehand, since for $T>0$ the
Euclidean $O(4)$ invariance is violated.

To conclude this section, one comment on nondiagonal terms in
(\ref{3}), which contribute to the total Hamiltonian $\hat H$ as
\cite{26} \be \hat H = H_0 (\mu_1, \mu_2) + V_{SD}^{(diag)} \hat
1_1 \hat 1_2 + V_{SD}^{(nond)} (\gamma_5)_1
(\gamma_5)_2+...\label{14}\ee where dots imply terms proportional
to $\hat 1_1(\gamma_5)_2$ and $(\gamma_5)_1 \hat 1_2$. It is clear
that $V_{SD}^{(nond)}$ has the structure similar to (\ref{4}) with
replacement $V_i\to V_i^{(nond)}$ and  contains  terms
proportional to electric correlators Eq.(\ref{5}). E.g. in
\cite{26} the term $V_4^{(nond)}$ was found to be \be V_4^{(nond)}
(R) =\int^\infty_{-\infty} d\nu \left(3D^E +3D_1^E+(3\nu^2
+R^2)\frac{\partial D^E_1}{\partial R^2}\right)\label{15}\ee For
heavy quarks (and for light quarks in the states, where SDI is
repulsive) the extremal values $\mu_i^{(0)}$ can be found from the
minimum of the spinless Hamiltonian $H_0(\mu_1, \mu_2)$, and in
this case SDI gives  spin corrections, which are not large even
for light quarks and are in good agreement   with experiment --
see \cite{18,19} for heavy and light quarkonia respectively.

For light quarks in the states with attractive spin-spin
interaction the $q\bar q$ Hamiltonian should be taken with the
full matrix structure as in (\ref{14}), and the stationary values
of the quark "constituent" masses $\mu_i^{(0)}$ should be found
from the minimum of eigenvalues of $\hat H$, see \cite{26} for
more details. In what follows we shall consider only diagonal part
of SDI, which is essential for heavy quarkonia.

\section{Spin interactions in the deconfinement phase}

In the confined phase SDI in terms of field correlators was
studied in \cite{11,12,15}$^-$\cite{19}. For $T>T_c$ the
correlator $D^E(x)$ vanishes, as it was argued in \cite{21,22,23}
and confirmed in lattice calculations \cite{24}. This fact leads
to a serious change both in SDI as well as in the spin-independent
part $H_0$ of the total Hamiltonian $\hat H$. The latter change
was studied in detail in recent papers \cite{27,28}, where it was
shown that
 in the  total static potential $V(R) =V_D(R) +V_1(R)$, generated
 by $D^E(x)$ and $D_1^E(x)$  respectively, only the term $V_1(R)$
 survives for $T>T_c$. This term in contrast to the confining
 potential $V_D(R)\sim \sigma R,~~ R\to \infty$, saturates at
 large $R$ and can support bound states of $c\bar c$ and $b\bar b$
 in some interval of temperatures $T_c\leq T\leq T_d$, while bound
 states dissociate at $T\geq T_d$. ($T_d\sim 2T_c$ for $D_1^E$
 calculated analytically in \cite{29}, and $T_d\sim 1.12 T_c$ for
 $D_1^E$ from lattice  calculations done in \cite{24}, see
 \cite{28} for details). All this dynamics is due to colorelectric
 correlator $D_1^E$ and colormagnetic fields were not taken into
 account. We now consider the role of SDI, which is mostly due to
 colormagnetic fields.

 We start with small distances and assume, according to
 \cite{27,28,29} the following form of  $D_1^E, D_1^H, D^H$
 \be
 D_1^{E,H} = D_1^{E,H} (pert)  + D_1^{E,H}(nonpert), ~~ D^H =
 D^H_{nonpert} +D^H_{pert},\label{16}\ee
 where
 \be D_1^{E,H}(pert) = \frac{16\alpha_s}{3\pi x^4}
 e^{-\lambda_{E,H} x} +O(\alpha_s^2);~~ D_1^{E,H}(nonpert, M_{E,H} x\gg1)=C_{E,H}
 \frac{e^{-M_{E,H}x}}{x}\label{17}\ee
and \be D^H_{nonpert} (xM_0\gg1) =d_H e^{-M_0 x}, ~~ D^H_{pert} =
\frac{g^4}{4\pi^4 x^4}    + O(\alpha_s^3)\label{18}\ee Here
$C_{E,H}, d_H$ are known   constants  depending on $\sigma_H$  -
the spacial string tension, $\sigma_H=\frac12 \int
D^H_{nonpert}(x) d^2 x$.

One can easily check in (\ref{8}-\ref{12}), that the small
distance behaviour of SDI potentials $V_i(R)$ does not change
much, since it is defined by the perturbative parts of correlators
which are little modified when temperature grows above $T_c$. The
main difference comes at large distances, where according to Eqs.
(\ref{8}-\ref{12}) one finds in the limit of large $R$,

\be V'_1(R) =- \sigma_H,~~ V'_2(R) = \frac{\gamma_H}{R}, ~~
\gamma_H=\int^\infty_{-\infty} d\nu \int^\infty_0 \lambda d\lambda
D^H_{nonpert} (\lambda, \nu)\label{19}\ee \be \varepsilon' (R)
=\frac12\int^\infty_{-\infty} d\nu RD_1^E(R) \to 0, ~~ V_3 (R) \to
0, ~~ V_4 (R) \to 0\label{20}\ee where in (\ref{20}) all three
potentials are exponentially small at large $R$. Therefore the
spin-orbit part of $V_{SD}^{(diag)}$ has the form (for equal quark
pole masses) \be V_{SD}^{diag} (R\to \infty) =- \frac{\veS
\veL\sigma_H}{\mu^2 R}+ \frac{\veS \veL \gamma_H}{\mu^2 R^2}
+O(e^{-MR}) .\label{21}\ee The first term on the r.h.s. of
(\ref{21}) was quoted (without derivation) in \cite{21}, where it
was suggested that being dominant at large $R$ (where $V_1(R)$
exponentially approaches to the  constant limit $V_1(\infty)$)
this term by itself can support bound states of heavy (and,
possibly, light) quarkonia. Indeed, considering  the first term in
(\ref{21}) as a Coulomb-like potential one arrives for heavy
quarkonia (where $\mu_0\approx m$) to the mass spectrum of bound
states for $\veS \veL >0, J=L+1$, \be M_n =2m
-\frac{\sigma_H^2}{4m^3} \frac{L^2}{(L+n+1)^2}, ~~
n=0,1,2,...\label{22}\ee

We conclude this section with discussion of hyperfine interaction
in the deconfined phase. As it is seen in (\ref{11}), $V_4(R)$
depends only on $D^H, D_1^H$ and does not change across $ T_c$
according to the lattice data \cite{24}. Keeping only (the
dominant) perturbative part of $V_4$, one obtains \be V_{hf}
=\frac{8\pi\alpha_s \vesig^{(1)} \vesig^{(2)}}{9\mu_1 \mu_2}
\delta^{(3)} (\veR).\label{23}\ee

For the hyperfine energy shifts this gives \be \Delta E_{hf}
=\frac{4\alpha_s \lan V'(R)\ran }{9(\mu_1+\mu_2)}
\left(\begin{array}{ll} -3,&S=0\\
+1,&S=1\end{array} \right).\label{24}\ee

To calculate $\lan V'(R)\ran$, one can use the static potential,
computed in \cite{27}, \cite{28} both analytically and on the
lattice.  A rough estimate can be found from   comparison of free
static energies found on the lattice (see e.g. \cite{30} and refs.
therein)  both below and above $T_c$ (up to $T\approx 1.3 T_c)$,
which shows a close similarity of $V'(R)$ in both temperature
domains. Hence one  expects for $T_c\leq T \leq 1.3 T_c$ the mass
gap between $J/\psi$ and $\eta)c$ of the same order as for  $T=0$,
i.e. $\delta E \sim 0.1$ GeV, which roughly agrees with lattice
MEM computations \cite{30}.

The situation with  hyperfine spin splittings in the light $q\bar
q$ system is however different and cannot be treated by the
methods given above, since the restoration of chiral symmetry at
$T\geq T_c$ needs the effective Lagrangian technic \cite{20}
mentioned above. The physical reason for that lies in the fact,
that for light quarks  $\Delta E_{hf}$ Eq. (\ref{24}),  becomes
dominant (for $\mu_1=\mu_2\approx m$) and cannot be treated as a
perturbation, and one needs to solve nonlinear equations for the
effective mass operator, given in \cite{20}. This point will be
treated elsewhere.

In summary, we have derived the spin-dependent potentials for the
$q\bar q$ system for any temperature $T$ valid in the situation,
when spin splittings can be considered perturbatively.

For 1.3 $T_c\geq T\geq T_c$ spin splittings of charmonium and
bottomonium are shown to change little compared to zero
temperature case. The color magnetic confinement produces the
Thomas spin-orbit term, which dominates at large distances, in
absence of colorelectric confinement  at $T\geq T_c$, and  can
possibly support a sequence of bound states. These features
demonstrate the importance of strong interaction in the
quark-gluon plasma, which was advocated in \cite{21}-\cite{24} and
supported by explicit calculations in \cite{27,28} in agreement
with lattice data \cite{30}.

This work is supported by the Federal  Program of the Russian
Ministry of industry, Science and Technology
  No.40.052.1.1.1112, and by the
grant for scientific schools NS-1774. 2003.


\begin{thebibliography}{99}
  \bibitem{1}
  A.De Rujula, H.Georgi and S.L.Glashow, Phys. Rev. {\bf D12}, 147
  (1975);\\
  T.de Grand, P.L.Jaffe, K.Johnson and J.Kiskis, Phys. Rev. {\bf D12}, 2060 (1975).

  \bibitem{2}
 T.Appelquist, R.M.Barnett and
K.D.Lane, Ann.  Rev.  Nucl. Part. Sci.  {\bf 28},  387 (1978);\\
W.Buchmueller, Phys.  Lett.  {\bf 112}, 479 (1982).

  \bibitem{3}
  E.Eichten, F.Feinberg, Phys. Rev. {\bf D23}, 2724 (1981).

  \bibitem{4} D.Gromes, Z.Phys. {\bf C26}, 401 (1984).

  \bibitem{5} W.Buchm\'{u}ller, Y.J.Ng and S.-H.H.Tye, Phys. Rev.
  {\bf D24}, 3003 (1981);\\
  J.Pantaleone, Y. J.Ng and S.-H.H.Tye, Phys. Rev.
  {\bf D33}, 777 (1986).


\bibitem{6}M.Campostrini,
Nucl.  Phys.{\bf B256},  717 (1985);\\ C.Michael and P.E.L.Rakow,
Nucl. Phys.  {\bf B256},  640 (1985);\\ P.de Forcrand and
J.D.Stack, Phys. Rev.  Lett.  {\bf 55},   1254 (1985);\\
M.Campostrini, K.Moriarty and
C.Rebbi, Phys.  Rev. Lett.  {\bf 57}, 44 (1986);\\
C.Michael, Phys.  Rev.  Lett. {\bf 56},  1219 (1986);\\ I.J.Ford,
J.Phys. G {\bf 15}, 1571 (1989);\\ A.Huntley and C.Michael,  Nucl.
Phys. {\bf B270},  123 (1986).

\bibitem{7} K.D.Born et al.  Phys.Lett.
{\bf B329}, 332 (1994);\\ G.S.Bali, K.Schilling and a.Wachter,
Phys. Rev. {\bf D55}, 5309 (1997), ibid. {\bf D56}, 2566 (1997).
\bibitem{8}
 M.Koma, Y.Koma, H.Wittig, hep-lat/0510059.


\bibitem{9} H.G.Dosch, Phys. Lett. B {\bf 190}, 177 (1987) ;\\
  H.G.Dosch and Yu.A.Simonov, Phys. Lett. B {\bf 205}, 339 (1988) ;\\
   Yu.A.Simonov, Nucl.  Phys.  B {\bf 307}, 512 (1988).

\bibitem{10}
 A.Di Giacomo, H.G.Dosch, V.I.Shevchenko, Yu.A.Simonov, Phys. Rept. {\bf 372}, 319 (2002);
 hep-ph/0007223.


\bibitem{11} Yu.A.Simonov, Nucl. Phys. {\bf B324}, 67 (1989).


\bibitem{12} M.Schiestl, H.G.Dosch, Phys. Lett. {\bf  B209}, 85
(1988).

\bibitem{13} A.Di Giacomo and H.Panagopoulos, Phys. Lett. {\bf B285},  133 (1992) ;\\
 M.D'Elia, A.Di Giacomo, and   E.Meggiolaro, Phys. Lett. {\bf
 B408},  315 (1997) ;\\
 A.Di Giacomo, E.Meggiolaro  and H.Panagopoulos,
 Nucl. Phys. {\bf B483},  371 (1997)\\
 M.D'Elia, A.Di Giacomo, and   E.Meggiolaro, Phys. Rev. {\bf
 D67},   114504 (2003);\\
 G.S.Bali, N.Brambilla, A.Vairo, Phys. Lett. {\bf B 42}, 265
 (1998);\\ E.Meggiolaro, Phys. Lett. {\bf B451}, 414 (1999) .


\bibitem{14}  A.M.Badalian, V.P.Yurov, Yad. Fiz. {\bf 56}, 239 (1993).

\bibitem{15} A.M.Badalian, Yu. A.Simonov, Phys. At. Nucl. {\bf 59}, 2164
(1996).

  \bibitem{16}
     Yu.A.Simonov, QCD and Topics in Hadron Physics, Lectures at the
       XVII International School of Physics, Lisbon, 29 September -4
       October,1999, hep-ph/9911237;\\
 Yu.A.Simonov, Spin-interactions of light quarks,
preprint ITEP 97-89 (unpublished).

\bibitem{17} Yu.A.Simonov, Phys. Rev. {\bf D65}, 116004 (2002).

\bibitem{18} A.M.Badalian, Phys. At. Nucl. {\bf 66}, 1342
(2003);\\ A.M.Badalian, B.L.G.bakker, Phys. Rev. {\bf D67}, 071901
(2003).
\bibitem{19} A.M.Badalian, V.L.Morgunov, B.L.G.Bakker, Phys. At.
Nucl. {\bf 63}, 1635 (2000).
\bibitem{20}
Yu.A.Simonov, Phys. Rev. {\bf D65}, 094018 (2002);\\
S.M.Fedorov, Yu.A.Simonov, JETP Lett. {\bf 78},  57 (2003).

\bibitem{21} Yu.A.Simonov, JETP Lett. {\bf 54}, 249 (1991).

\bibitem{22} Yu.A.Simonov, JETP Lett. {\bf 55}, 605 (1992).

\bibitem{23} Yu.A.Simonov, Phys. At. Nucl. {\bf 58}, 309 (1995);
hep-ph/9311216;\\
N.O.Agasian, JETP Lett. {\bf 57}, 208 (1993).

\bibitem{24} M. D'Elia, A. Di Giacomo and E. Meggiolaro, Phys. Rev. D {\bf 67}, 114504
(2003);\\
A. Di Giacomo, E. Meggiolaro and H. Panagopoulos, Nucl. Phys. B
{\bf 483}, 371
(1997);\\
A. Di Giacomo, E. Meggiolaro and H. Panagopoulos,
hep--lat/9603018.
\bibitem{25}
Yu.A.Simonov, J.A. Tjon, Ann. Phys.  {\bf 300}, 54 (2002);
hep-ph/0205165;\\ Yu.A.Simonov and J.A.Tjon, Ann Phys. {\bf 228},
1 (1993).

\bibitem{26}Yu.A.Simonov, Phys. At. Nucl. {\bf 68}, 709 (2005).


\bibitem{27} Yu.A.Simonov, Phys. Lett. {\bf B619}, 293 (2005),
hep-ph/0502078.
\bibitem{28} A.Di Giacomo, E.Meggiolaro, Yu.A.Simonov,
A.I.Veselov, hep-ph/0512125.
\bibitem{29}Yu.A.Simonov, Phys. At. Nucl. {\bf 69}, N3,
(2006), hep-ph/0501182.


\bibitem{30}P.Petreczky, hep-lat/0409139.


  \end{thebibliography}
  \end{document}